\documentclass[sigconf]{acmart} 

\usepackage{dsfont}
\usepackage{amsmath}

\providecommand{\lin}[1]{\ensuremath{\left\langle #1 \right\rangle}}
\providecommand{\abs}[1]{\ensuremath{\left\lvert#1\right\rvert}}

  \providecommand{\R}{\mathbb{R}} 
  
  \DeclareMathOperator{\E}{{\mathbb E}}
  \providecommand{\EEb}[2]{\ensuremath{\E_{#1}\!\! \left[#2\right] }} 
  \providecommand{\prob}[1]{\ensuremath{\mathbb{P}\left[#1\right] } }

  \newcommand\independent{\protect\mathpalette{\protect\independenT}{\perp}}
  \def\independenT#1#2{\mathrel{\rlap{$#1#2$}\mkern2mu{#1#2}}}

  \DeclareMathOperator*{\argmin}{arg\,min}

  \providecommand{\1}{\mathbf{1}}

  \providecommand{\cc}{\mathbf{c}}
  \providecommand{\dd}{\mathbf{d}}
  \providecommand{\ee}{\mathbf{e}}

  \providecommand{\qq}{\mathbf{q}}
  \providecommand{\rr}{\mathbf{r}}

  \providecommand{\ww}{\mathbf{w}}
  \providecommand{\xx}{\mathbf{x}}

  \providecommand{\cD}{\mathcal{D}}
  \providecommand{\cE}{\mathcal{E}}
  \providecommand{\cF}{\mathcal{F}}
  \providecommand{\cG}{\mathcal{G}}

  \providecommand{\cL}{\mathcal{L}}
  
  \providecommand{\cN}{\mathcal{N}}

  \providecommand{\cR}{\mathcal{R}}

\AtBeginDocument{%
  \providecommand\BibTeX{{%
    \normalfont B\kern-0.5em{\scshape i\kern-0.25em b}\kern-0.8em\TeX}}}

\setcopyright{acmcopyright}
\copyrightyear{2021}
\acmYear{2021}
\acmDOI{}

\acmConference[]{}{}{}
\acmBooktitle{}
\acmPrice{}
\acmISBN{}

\usepackage{balance}

\usepackage{algorithm}
\usepackage{algpseudocode}

\usepackage{subcaption}

\usepackage[capitalize,noabbrev]{cleveref}

\usepackage[shortlabels]{enumitem}

\usepackage[acronym, nomain, nopostdot, nonumberlist, toc]{glossaries}
\newacronym{ltr}{LTR}{learning-to-rank}
\newacronym{pbm}{PBM}{position bias model}
\newacronym{cpbm}{CPBM}{contextual position bias model}
\newacronym{ctr}{CTR}{click-through rate}
\newacronym{ubm}{UBM}{user browsing model}
\newacronym{dcm}{DCM}{dependent click model}
\newacronym{em}{EM}{expectation-maximization}
\newacronym{ih}{IH}{intervention harvesting}
\newacronym{ir}{IR}{information retrieval}
\newacronym{mle}{MLE}{maximum likelihood estimator}
\newacronym{dcg}{DCG}{discounted cumulative gain}
\newacronym{rbp}{RBP}{rank-biased precision}

\begin{document}
\title{Ranker-agnostic Contextual Position Bias Estimation}


\settopmatter{authorsperrow=3}

\author{Oriol Barbany Mayor}
\email{oriol.barbanymayor@alumni.epfl.ch}
\authornote{Work done during an internship at Amazon Music ML.}
\affiliation{%
	\institution{EPFL}
	\city{Lausanne}
}
\author{Vito Bellini}
\email{vitob@amazon.com}
\affiliation{%
	\institution{Amazon Music ML}
	\city{Berlin}
}
\author{Alexander Buchholz}
\email{buchhola@amazon.com}
\affiliation{%
	\institution{Amazon Music ML}
	\city{Berlin}
}
\author{Giuseppe Di Benedetto}
\email{bgiusep@amazon.com}
\affiliation{%
	\institution{Amazon Music ML}
	\city{Berlin}
}
\author{Diego Marco Granziol}
\authornotemark[1]
\email{diego@robots.ox.ac.uk}
\affiliation{%
	\institution{Machine Learning Research Group}
	\city{University of Oxford}
}
\author{Matteo Ruffini}
\email{ruffinim@amazon.com}
\affiliation{%
	\institution{Amazon Music ML}
	\city{Berlin}
}
\author{Yannik Stein}
\email{syannik@amazon.com}
\affiliation{%
	\institution{Amazon Music ML}
	\city{Berlin}
}

\renewcommand{\shortauthors}{Barbany Mayor, et al.}

\begin{abstract}
	\Gls*{ltr} algorithms are ubiquitous and necessary to explore the extensive catalogs of media providers. To avoid the user examining all the results, its preferences are used to provide a subset of relatively small size. The user preferences can be inferred from the interactions with the presented content if explicit ratings are unavailable. However, directly using implicit feedback can lead to learning wrong relevance models and is known as biased \gls*{ltr}. The mismatch between implicit feedback and true relevances is due to various nuisances, with position bias one of the most relevant. Position bias models consider that the lack of interaction with a presented item is not only attributed to the item being irrelevant but because the item was not examined. This paper introduces a method for modeling the probability of an item being seen in different contexts, e.g., for different users, with a single estimator. Our suggested method, denoted as contextual \gls*{em}-based regression, is ranker-agnostic and able to correctly learn the latent examination probabilities while only using implicit feedback. Our empirical results indicate that the method introduced in this paper outperforms other existing position bias estimators in terms of relative error when the examination probability varies across queries. Moreover, the estimated values provide a ranking performance boost when used to debias the implicit ranking data even if there is no context dependency on the examination probabilities.
\end{abstract}



\maketitle

\glsresetall

\section{Introduction}
\Gls*{ltr} algorithms are essential for presenting a reduced version of the otherwise overwhelming content catalog to consumers. The ultimate goal of these algorithms is to learn the relevance of each item in the catalog according to the user preferences to provide an improved experience and a better engagement. In most real scenarios, explicit ratings for each item do not exist or are scarce, which means that the true preferences of the users are typically unknown. To overcome the lack of explicit ratings, interaction metrics such as the clicks or the dwell time are recorded. These interaction metrics, also known as implicit feedback, are then used to learn the preference profiles \cite{unbiased_ltr, unbiased_lambdamart}.

It is well-known that implicit feedback suffers from various type of biases due to both the system and the user. Examples of such biases are position bias, trust bias, quality bias, and context bias \cite{pb_experiment, other_biases}. This paper focuses on modeling the position bias, the major type of bias among the latter \cite{unbiased_lambdamart}. The position bias models the fact that the user typically does not examine all presented items and is more inclined to engage with top-ranked ones. This means that lower-ranked items may not be clicked even if they are more relevant to the user than higher-ranked ones. This leads to top-ranked items being more likely to collect feedback, which in turn can influence future rankings and promote misleading rich-get-richer dynamics \cite{unbiased_ltr,pb_experiment}. The position bias effect is exacerbated when the results are presented using swipeable carousels. In this case, not all the results are shown at the same time, which makes the items presented in the last positions less likely to be examined.

The position bias is formalized in many click models such as the \gls*{ctr} model, the cascade model, the \gls*{ubm}, and the \gls*{dcm} \cite{click_models, cascade}. In this project, we model the position bias with the \gls*{cpbm} \cite{contextual_ih}, a generalization of the \gls*{pbm} \cite{pbm}. The \gls*{pbm} is a click model where the click event depends on both the user examining the item and on its relevance. The estimated position bias can be then used for debiasing the implicit data before using it to learn the user preferences, improving the performance of \gls*{ltr} models \cite{unbiased_lambdamart, unbiased_ltr, unbiased_ltr_propensity}.
In fact, as proven by \cite{unbiased_ltr}, given the correct position bias estimates, the unbiased rankers converge to the models trained with explicit ratings.

In this paper, we empirically show that the estimates obtained under the \gls*{cpbm} are better at debiasing the data compared to those obtained following the \gls*{pbm}. This translates into an improved ranking performance. Unlike other existing methods, the proposed estimator is ranker-agnostic and can learn the position bias from implicit feedback collected without interventions that can harm the user experience.

The rest of this paper is structured as follows. \Cref{sec:setting} presents the \gls*{ltr} setting and formalizes the \gls*{pbm}. This section also introduces the \gls*{cpbm} and motivates its need. In \cref{sec:related_work} we review the literature on position bias estimators. In particular, we discuss the methods that follow the \gls*{pbm} and \gls*{cpbm}. \Cref{sec:methodology} introduces our contribution: a novel method for \gls*{cpbm} estimation inspired by the regression-based \gls*{em} algorithm described in \cite{regression_em}. The experimentation setup and the discussion of results are included in \cref{sec:exp}. Finally, we conclude in \cref{sec:conclusions}.

\section{Setting}
\label{sec:setting}

Let $\qq\ \overset{\text{i.i.d.}}{\sim} Q$ be the context vector issued for each query, which can contain information, among others, about the user's device, the query itself, and the user's preferences. Let $\Omega(\qq)$ be the subset of items in the catalog that are potentially relevant to $\qq$, where each element $\dd\in\Omega(\qq)$ is a vector representing the features of one item of the catalog. For each context $\qq$, a ranker presents a $K-$permutation of the subset $\Omega(\qq)$ to the user. We will denote the ranking presented to the user as $\cD:=(\dd^1,\dots,\dd^K)$. The interactions with the user, along with the context vector and the presented ranking, are recorded and referred to as implicit feedback.

The implicit feedback considered in this paper are click events, which are modeled by a Bernoulli random variable $C$. Following the \gls*{pbm} \cite{pbm}, the probability that the user clicks on an item $\dd\in\Omega(\qq)$ for $\qq\overset{\text{i.i.d.}}{\sim} Q$ presented at position $k\in[K]:=\{1,2,\dots,K\}$, can be factored as follows:
\begin{align}
	\prob{C=1|\qq,\dd,k}=\prob{E=1|k}\prob{R=1|\qq,\dd}\,,
	\label{eq:pbm}
\end{align}
where $E,R$ are latent Bernoulli random variables representing the event of the user observing an item at position $k$ and the relevance of item $\dd$ given a context $\qq$, respectively.

We are interested in modeling the probability of a user observing an item at a given position. Note from \eqref{eq:pbm} that the \gls*{pbm} assumes that $(E \independent \qq,\dd) | k$. In other words, an item displayed in a given position is examined regardless of the context and the item itself. It is safe to assume that $E \independent \dd | k$ because the item cannot influence the probability of examination if it is unseen. However, there are some cases where the examination event may depend on the context:

\begin{enumerate}[(a)]
	\item \label{it:queries} Navigational (specific queries targeting a particular item) vs. Informational queries (broad queries with numerous relevant results) as described in \cite{contextual_ih}. In the former only a few results will be observed, while in the latter more results are prone to being examined.
	\item \label{it:devices} Different number of visible items per device, e.g., swipeable carousel with more items in the desktop version than on the smartphone version. In this case, even the same user will probably examine more results when exploring the catalog from a computer than from a cellphone.
	\item \label{it:user} User-specific browsing patterns. Some users might be satisfied with the first results, while some others might prefer to explore more thoroughly the presented items.
\end{enumerate}

This paper uses the \gls*{cpbm} described in \cite{contextual_ih}, which assumes that the click probability can be factored as follows
\begin{align}
	\prob{C=1|\qq,\dd,k}=\prob{E=1|\qq,k}\prob{R=1|\qq,\dd}\,.
	\label{eq:cpbm}
\end{align}

Let $\cL:=\{(\cc^i,\qq^i, \cD^i)\}_{i\in[N]}$ be the click log from which we want to learn the contextual position bias, where $\cc:=[\cc_1,\dots,\cc_K]^T\in \{0,1\}^K$ is the vector of observed click realizations obtained after presenting $\cD$ to the user. In particular, $\cc_k$ is an indicator of the click received for item $\dd^k\in \cD$. The likelihood of observing the clicks in $\cL$, assuming that clicks are independent, can be written as

\begin{align}
	\prod_{(\cc,\qq,\cD)\in \cL} \prod_{k\in [K]} \prob{C=1|\qq,\dd^k,k}^{\cc_k}\bigg(1-\prob{C=1|\qq,\dd^k,k}\bigg)^{1-\cc_k}\,.
	\label{eq:likelihood}
\end{align}

\section{Related Work}
\label{sec:related_work}
This section describes some existing position bias estimators that are relevant for this work. The estimators are classified into contextual and non-contextual, which model the click probability using \eqref{eq:pbm} and \eqref{eq:cpbm}, respectively.

\subsection{Position bias estimators}
\label{sec:pbm_estim}
One of the simplest click models that take into account the influence of the ranking position is the rank-based \gls*{ctr} model \cite{click_models}. This model assumes that an item is clicked with a probability that uniquely depends on the ranking position. Put differently, this model assumes that $\prob{C=1|\qq,\dd,k}=\prob{C=1|k}$. The \gls*{mle} for this model is the proportion of clicks per position \cite{click_models}. One can normalize such quantities by the estimated value at the first position and obtain a naive yet decent estimator of the position bias (see results in \cref{sec:exp}). However, this estimator does not take the relevance of an item into account. That is, the influence of the latent variables $R$ on the clicks, is not considered in the \gls*{ctr} model, which is overly simplistic.

The \gls*{mle} of the \gls*{pbm} is obtained by maximizing \eqref{eq:likelihood} with respect to the latent parameters of the relevance and examination models after replacing the click probability by \eqref{eq:pbm}. The drawback of this approach is that it involves modeling the latent relevance model, which is as hard as solving the \gls*{ltr} problem \cite{contextual_ih}. 

Estimating the position bias with this approach is still possible, and can be done using the closed form updates of the \gls*{em} algorithm \cite{regression_em}. The maximization step of the \gls*{em} estimator requires aggregating the clicks obtained at the same position $k$ and the same $(\qq,\dd)$ pair to compute the examination and the relevance models, respectively. However, $\qq$ is rarely repeated across $\cL$ in practice. For this reason, \cite{regression_em} proposed to pose the relevance model estimation as a regression problem and only use the closed-form maximization update for the examination model. This algorithm does not require a model of the ranker, but lacks provable guarantees of global optimality similarly to standard \gls*{em} \cite{em_convergence} and offers poor performance in some setups \cite{ih}.

One simple method to compute the position bias without modeling the relevance, is to apply a controlled randomization by swapping the order of the presented items \cite{unbiased_ltr}. The estimator obtained using such swapping interventions has provable theoretical guarantees and offers better performance compared to the regression-based \gls*{em} method \cite{ih}. However, this is an intrusive method that harms user experience due to the randomization of the ranking.

Different randomization strategies have been proposed. The strategy introduced in \cite{unbiased_ltr} requires swapping the top position with every other ranking position. In order to avoid displaying low ranked items in the first positions, \cite{regression_em} proposed to perform swaps between items $(k, k+1)$ for each $k\in[K-1]$ separately. This modification gives similar performance to the former method, while dampening the decrease in user engagement \cite{regression_em}. One can also randomly allow either swapping an item at an odd position with the next item, or swapping an item at an even position with the following item. Using the same computations as in \cite{ih,regression_em}, it is easy to see that the the position bias curves computed with this last randomization strategy are statistically consistent estimates of the relative position bias as $N$ grows \cite{ih}. Given that the last randomization strategy makes a better use of the data, we implemented this last approach. In fact, using the chosen randomization strategy allows using half of the data in expectation to estimate the position bias at a given position instead of only using a $K-$th fraction as in the former randomization strategies. Therefore, the estimates will have lower variance, and thus the swap-based methods will constitute a stronger baseline.

In an attempt to avoid the previous ranking randomization while still not fully modeling the relevance, \cite{ih} proposed the \gls*{ih} estimator. In this approach, a ranker is chosen uniformly at random from a set of rankers and used to present the results to the user. With this method, only modeling an average model of the relevance suffices. Contrary to the ranking randomization needed for swap-based methods, this method does not degrade the overall ranking performance if the rankers are good enough. Moreover, this method makes more efficient use of data than the swap-based approach described in \cite{unbiased_ltr}, which translates to lower variance of the estimates \cite{ih}. The downside of this approach is that it needs to maintain a set of rankers, and they have to disagree enough. The discrepancy requirement is not prohibitive in practice \cite{ih}, but maintaining various rankers may be restrictive in some industrial applications. The reason is that, in this setting, each ranker only processes a portion of the data. Therefore, the amount of data needed to obtain the same performance for a given ranker increases linearly with the number of rankers. Moreover, using historical data that has been recorded with a single ranker or whose ranking process is unknown is not possible with \gls*{ih} methods. Overall, these make \gls*{ih}-based approaches unattractive.

\subsection{Contextual position bias estimators}
\label{sec:cpb_estimators}
Suppose the position bias is only affected by the types of query described in \ref{it:queries}, where there would be only two different position bias curves in the data. Here it may be a good option to filter the implicit data by query type and fit one position bias estimator using the methods discussed in \cref{sec:pbm_estim} on each subset of data. In the following, the methods where the data is filtered before feeding it to the estimators will be denoted with the semi-contextual prefix.

The same approach can be used in setting \ref{it:devices}, where the position bias is fully determined by the ranking position and the device used to perform the query. Nonetheless, in the case of having devices that are poorly represented on the data, the estimations will have large variance. Moreover, this approach would fail to generalize to a possibly unseen device. This problem is brought to the extreme with setting \ref{it:user}, where we have a position bias per user. Following a semi-contextual approach, one model has to be computed and stored for each user, which is impractical. Moreover, the estimates for users with few recorded iterations will have a large variance.

As pointed in \cite{contextual_ih}, one straight-forward solution to compute a context-dependent position bias is to use a generative-modeling approach. Following this technique, the likelihood of $\cL$ is jointly maximized over the choice of the estimators of the relevance and examination models. However, this approach needs a model of the relevance and gives poor performance in practice (see \cref{sec:exp}).

To tackle these problems, the \gls*{ih} approach can be adapted for \gls*{cpbm} estimation by using the context to condition the prediction of the average relevance and the position bias \cite{contextual_ih}. This approach potentially generalizes to poorly represented and even unseen contexts and offers the advantages of harvesting-based methods. However, the contextual \gls*{ih} method described in \cite{contextual_ih} has the same drawbacks as its non-contextual version. Namely, the contextual \gls*{ih} approach requires maintaining several rankers, each of them trained with only a portion of the data, and requires logging both all the rankings given by each ranker and an indicator of the presented ranking.

The method proposed in \cite{attributebasedpropensity} consists in a EM algorithm where the position bias parameters can depend on \emph{attributes}, such as device type. The main difference with our work is that in \cite{attributebasedpropensity} the position bias curves are independently estimated for each attributes' configuration, while we adopt a neural network to jointly model the position bias depending on a context vector.
Another related work is \cite{recommendingwhatvideo}, where a neural network is fed with contextual information to  estimate the position bias.

In \cref{sec:methodology} we present a method that, similarly to the former approaches, is able to model context-dependent latent examination probabilities. The proposed method is ranker-agnostic, meaning that it does not require using several rankers when collecting the data nor logging information about the ranking process. Moreover, the presented rankings have no need to be randomized. In conclusion, the proposed algorithm can be used in virtually any click log and avoids harming user experience in the data collection process.

\section{Methodology}
\label{sec:methodology}

In this section we present our contribution. When a black-box ranker was used to generate $\cL$ and no randomization was applied, the number of position bias estimators that we can use shrinks. In particular, among the estimators reviewed in \cref{sec:related_work}, only the \gls*{ctr} and \gls*{em} methods can work in this setup. Therefore, the \gls*{em}-based method proposed by \cite{regression_em} is arguably the best and most principled option under this framework. However, the latter does not allow to model the examination probabilities as a function of the context. With this in mind, we propose an extension of their method to the contextual case. \Cref{alg:em} presents an \gls*{em}-based contextual position bias estimator where the maximization steps for both the relevance and examination models are posed as a regression problem. Let $\widehat{\mathbb{P}}[E=1|\qq,k]$ and $\widehat{\mathbb{P}}[R=1|\qq,\dd]$ be the estimations of the mean of the examination and the relevance random variables, respectively. In \cref{alg:em}, we denote $f(\qq,k):=\widehat{\mathbb{P}}[E=1|\qq,k]$ and $g(\qq,\dd):=\widehat{\mathbb{P}}[R=1|\qq,\dd]$.

\begin{algorithm*}[t]
    \caption{\textsc{Contextual EM-Based Regression$(\cL)$}}
    \label{alg:em}
    \begin{algorithmic}[1]
    \State \textbf{Input: }Click log $\cL$, number of iterations $T$, class of estimators $\cF,\cG$, and loss $\ell(\cdot,\cdot)$
    \State \textbf{Output: }Estimated examination model $f^*$ and estimated relevance model $g^*$
    \State \textbf{Initialize: }$f_0\in\cF,g_0\in\cG$
    \For{$t=0,1,\dots,T-1$}
    	\For{mini-batch $\cL' \subseteq \cL$}
	        \State Let $\cL'_{f_{t}}:=\{(\cc,\qq,\cD, \cE:=[\ee_1,\dots, \ee_K] )\}: \ee_{k}\sim Ber(\widehat{\mathbb{P}}[E=1|\cc_k,\qq,\dd^k,k,f_t,g_t]) \ \forall (\cc,\qq,\cD)\in\cL'\}$
	        \State Let $\displaystyle f_{t+1} :=\argmin_{f\in\cF}\sum_{(\cc,\qq,\cD, \cE)\in \cL'_{f_{t}}} \sum_{k\in [K]} \ell(f(\qq,k),\ee_k)$
	        \State Let $\cL'_{g_{t}}:=\{(\cc,\qq,\cD, \cR:=[\rr_1, \dots, \rr_K])\} :\rr_k\sim Ber(\widehat{\mathbb{P}}[R=1|\cc_k,\qq,\dd^k,k,f_{t+1},g_t]) \ \forall (\cc,\qq,\cD)\in\cL'\}$
        	\State Let $\displaystyle g_{t+1}:=\argmin_{g\in\cG}\sum_{(\cc,\qq,\cD,\cR)\in \cL'_{g_{t}}} \sum_{k\in[K]} \ell(g(\qq,\dd^k),\rr_k)$
        \EndFor
    \EndFor
    \State \textbf{Return: }$f_T,g_T$
    \end{algorithmic}
\end{algorithm*}

The expectation step of \cref{alg:em} estimates the distribution of the latent variables $E,R$. Following the \gls*{cpbm}, we can easily compute the joint distribution of $(E,R)|C,\qq,\dd,k$ as detailed in \cite{regression_em}. By replacing the true distributions of the latent variables by their estimation given by $f,g$, we obtain the following estimated marginals:

\begin{align}
	\widehat{\mathbb{P}}[E=1|c,\qq,\dd,k,f,g]=c+(1-c)\frac{f(\qq,k)(1-g(\qq,\dd))}{1-f(\qq,k)g(\qq,\dd)}\,,
	\label{eq:marginal_e}
\end{align}

\begin{align}
	\widehat{\mathbb{P}}[R=1|c,\qq,\dd,k,f,g]=c+(1-c)\frac{(1-f(\qq,k))g(\qq,\dd)}{1-f(\qq,k)g(\qq,\dd)}\,.
	\label{eq:marginal_r}
\end{align}

The maximization step updates the parameters of $f,g$ using the quantities from the expectation step. In particular, we choose the hypothesis classes $\cF,\cG$ to be neural networks. In \cite{regression_em}, the regression problem of fitting the previous marginals, is transformed into a classification problem by sampling Bernoulli random variables from such marginals. Given that neural networks can easily be used as both classifiers and regressors, there is no need to pose the former regression problem as a classification problem. Therefore, this paper explores the performance of both \cref{alg:em} and a version of the latter where $f,g$ directly fit the marginals. This corresponds to setting $\ee_k=\widehat{\mathbb{P}}[E=1|\cc_k,\qq,\dd^k,k,f_t,g_t]$ on $\cL'_{f_{t}}$ (similarly for $\rr_k$ on $\cL'_{g_{t}}$) instead of setting it to the realizations of Bernoulli random variables parametrized by the marginals as in \cref{alg:em}.

Overall, the proposed method overcomes the limitations of swap- and \gls*{ih}-based algorithms as it requires no specific data collection method nor information about the ranking process. 
Unlike the original \gls*{em} estimator in \cite{regression_em}, the proposed method learns a context-dependent position-bias curve, which is more realistic (see examples \ref{it:queries}, \ref{it:devices}, and \ref{it:user}); it estimates the examination and relevance latent variables using neural networks trained in a mini-batches, instead of tree-based methods trained on the full dataset; it allows the option to treat the relevance prediction as a regression problem;
This results in more computationally and data efficient algorithms, which is especially suited for industrial settings. 
While context-dependent position-bias estimation can also be done with the contextual \gls*{ih}-based method described in \cite{contextual_ih}, the latter would require several rankers to be adopted, as aforementioned, with each of them only trained on a portion of the data. In \cref{sec:exp} we show that even when the click log $\cL$ is specifically collected to use the contextual \gls*{ih} method, our algorithm achieves better estimations in terms of the relative error.

\section{Experiments}
\label{sec:exp}

This section presents the experiments performed to compare the algorithms described in \cref{sec:related_work,sec:methodology}. We restrict training of all estimators to 50 epochs. With this constraint, using mini-batch updates consistently achieves better performance than using full batch updates for all the contextual estimators. For this reason, we avoid showing the results for full batch updates to improve the readability of the figures. We use a mini-batch size of 20 across all the experiments.

 The functions $\cF,\cG$ are both multi-layer perceptrons with a single hidden layer and sigmoidal activations. 
 $\cF$ takes the vector $\qq$ as input and outputs a $K$-dimensional vector representing the examination probability at each position conditioned on the context. The hidden size of $\cF$ is set to $2K$. $\cG$ is fed with the concatenation of the context and item feature vectors, and provides a single value representing the relevance of such item under the given context. We choose the hidden size to be half the input size. Tuning the hidden size hyper-parameter as well as using other functions for $\cF,\cG$ is left as future work.

 The estimators $f,g$ are tuned using empirical risk minimization with risk $\ell(\cdot,\cdot)$. We set $\ell$ to the binary cross-entropy loss or the mean squared error when $f,g$ solve a classification or regression task, respectively. The former will be referred as contextual \gls*{em}, and the latter is labeled as contextual \textsc{PEM}, which stands from probability-fitting \gls*{em}. 
 
 Solving the minimization problem in the maximization step of \cref{alg:em} for each of the latent variables at each iteration is prohibitively expensive depending on the stopping criteria. Given that the proposed estimator is targeted for industrial settings, we wanted to obtain fast estimations. With this in mind, we empirically found that using a single mini-batch gradient update works well. Thus, the results in this section are obtained by replacing the $\argmin$ operator in \cref{alg:em} by a single mini-batch gradient update. In particular, the model parameters are updated using ADAM with its default values \cite{adam}. 
 
For the \gls*{ih} method, we used the randomization strategy described in \cref{sec:related_work}. The items in positions $(k,k+1)$ are randomly swapped, where $k$ is either odd or even. Each individual swap between items $(k,k+1)$ is performed randomly. This implicitly yields several different rankings, where only one of them is revealed to the user. With this randomization, we calculate the intervention sets required for the \gls*{ih} method, see \cite{ih,contextual_ih}.

\subsection{Datasets}

The experiments are performed using two different datasets. In both cases, we set the contextual position bias as in \cite{contextual_ih}. That is,

\begin{align}
	\prob{E=1|\qq,k}=\frac{1}{k^{\max(\lin{\ww,\qq} + 1, 0)}}\,,
	\label{eq:true_cpb}
\end{align}
where $\ww:=\tilde{\ww}-\frac{1}{|\tilde{\ww}|}\lin{\tilde{\ww}, \1}$ is a zero mean random vector with $\tilde{\ww}_i \overset{\text{i.i.d.}}{\sim} U(-\eta,\eta) \ \forall i \in[\dim(\qq)]$. Note that the context dependency of the position bias increases with $\eta$, and the position bias solely depends on the position when $\eta=0$. The former ground-truth position bias is used in the generation of implicit feedback. The goal is to retrieve this quantity with the estimators introduced in \cref{sec:related_work,sec:methodology}.

For both datasets, we set the context dimension $\dim(\qq)=10$, and we collect clicks for the items presented in the first $K=10$ ranking positions.

To generate the click log $\cL$, we rank the items with an online \gls*{ltr} algorithm and record the received clicks. In particular, we use the \textsc{LinTSPBMRank} algorithm described in \cite{amazon_paper} with flat position bias. This algorithm is a version of Linear Thompson Sampling that considers the \gls*{pbm}. Setting the position bias to the all-ones vector corresponds to biased \gls*{ltr}, where clicks are used as relevances. In \cref{sec:unbiased_ltr} we assess the importance of the position bias in \textsc{LinTSPBMRank} by using the estimated position biases.

\subsubsection{SINBIN dataset} One of the datasets used in this work is the SINBIN dataset described in \cite{amazon_paper}. This dataset uses synthetic context and item feature vectors. The reward is computed using linear regression on the concatenation of context and item features. A binary version of such reward is issued if the item is seen according to the sampled examination random variables. This latter is the only difference with respect to the SINBIN implementation in \cite{amazon_paper}, where the obtained reward correspond to the binary relevance of an item divided by the probability of examining it. Note that retrieving the position bias with the rewards obtained using the original SINBIN dataset would be trivial, since the value of interest would be the reciprocal of the non-zero values obtained at each position. For this reason, our modification is more suited to test position bias estimators. We refer the interested reader to the original SINBIN paper for further details about the data generation process.

When using the SINBIN dataset, the position bias estimators use a click log containing 10,000 queries. For the unbiased \gls*{ltr} experiment in \cref{sec:unbiased_ltr}, 10,000 unseen queries are used. Since the number of items is relatively small $\Omega(\qq)$ is taken to be the set of all possible items.

\subsubsection{Contextual Yahoo LTR dataset} The Yahoo \gls*{ltr} Challenge corpus \cite{yahooltr} is widely used as a reference dataset for the \gls*{ltr} task \cite{deepprop, contextual_ih, unbiased_ltr, unbiased_lambdamart, ih}. This corpus contains several queries that are judged using a scale from 0 to 4. Since we are interested in using click data, we binarize the former relevances using the same procedure as in \cite{unbiased_ltr}. The original dataset has no context vectors, which are required for the proposed method. In order to generate them for each query, we follow a slight variation of the procedure described in \cite{contextual_ih}. In summary, the context vectors are generated as follows:

\begin{itemize}
	\item Rank the items optimally according to their true relevances. 
	\item Predict the previously obtained rewards at positions $k\in[K]$ with logistic model $r^k$. Such models take the average of the item feature vectors for each query as input and learn the parameter $\ww^k$ in $r^k(\xx) := \frac{1}{1+e^{-\lin{\ww^k, \xx}}}$.
	\item Let $s^j := max_m \abs{\ww^m_j}$ and select 5 features at random among the 30 features with largest $s^j$.
	\item Let the first half of the context vector $\qq\in\R^{10}$ be the value of the previous features on the average of the item feature vectors whose binary relevance is 1. The second half of $\qq$ is formed by i.i.d. samples from $\cN(0,\sigma^2)$.
\end{itemize}

When using this dataset, the position bias estimators use a click log obtained with the training partition of the set number 1 of the Yahoo \gls*{ltr} Challenge corpus. After filtering out the queries without relevant items as in \cite{contextual_ih}, this dataset contains 9,554 queries. The test partition of the same set, amounting to 3,397 queries after the filtering, is used for the experiments in \cref{sec:unbiased_ltr}. For more details about the contextual Yahoo \gls*{ltr} dataset, please refer to \cite{contextual_ih}.

\begin{figure*}
	\centering
	\begin{subfigure}{\columnwidth}
		\centering
		\includegraphics[width=\textwidth]{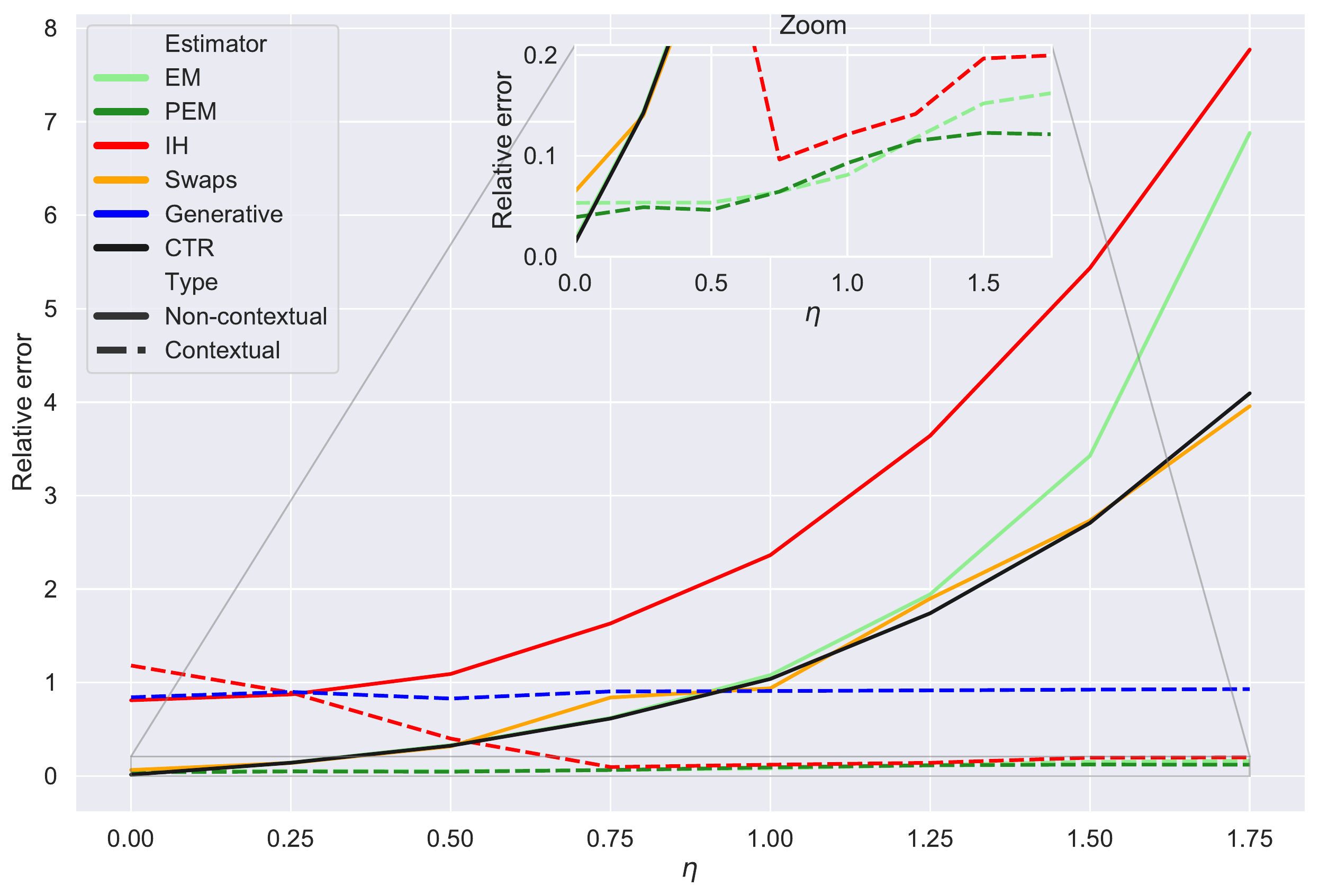}
		\caption{SINBIN dataset}
		\Description{The proposed method achieves the best performance. As expected, non-contextual methods have worse performance as the eta parameter increases.}
		\label{fig:relerror-eta-sinbin}
	\end{subfigure}%
	\begin{subfigure}{\columnwidth}
		\centering
		\includegraphics[width=\textwidth]{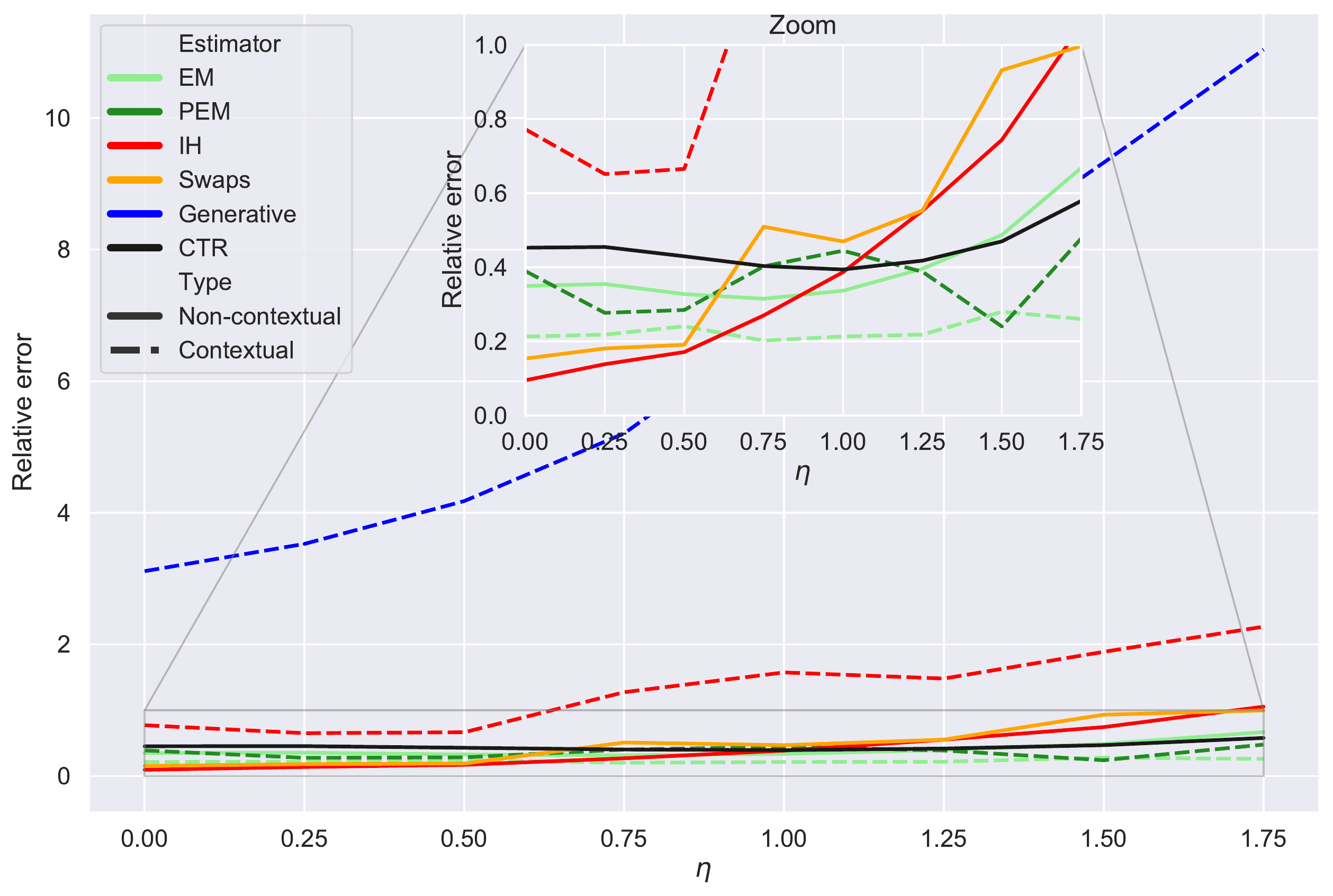}
		\caption{Contextual Yahoo \gls*{ltr} dataset}
		\Description{In this case, the proposed method and contextual IH method are the best. However, the contextual IH method offers poor performance when data is randomized, thus showing lack of robustness.}
		\label{fig:relerror-eta-yahoo}
	\end{subfigure}
	\caption{Empirical relative error as a function of the $\eta$ parameter.}
\end{figure*}%
\begin{figure}[t]
	\centering
	\includegraphics[width=.5\textwidth]{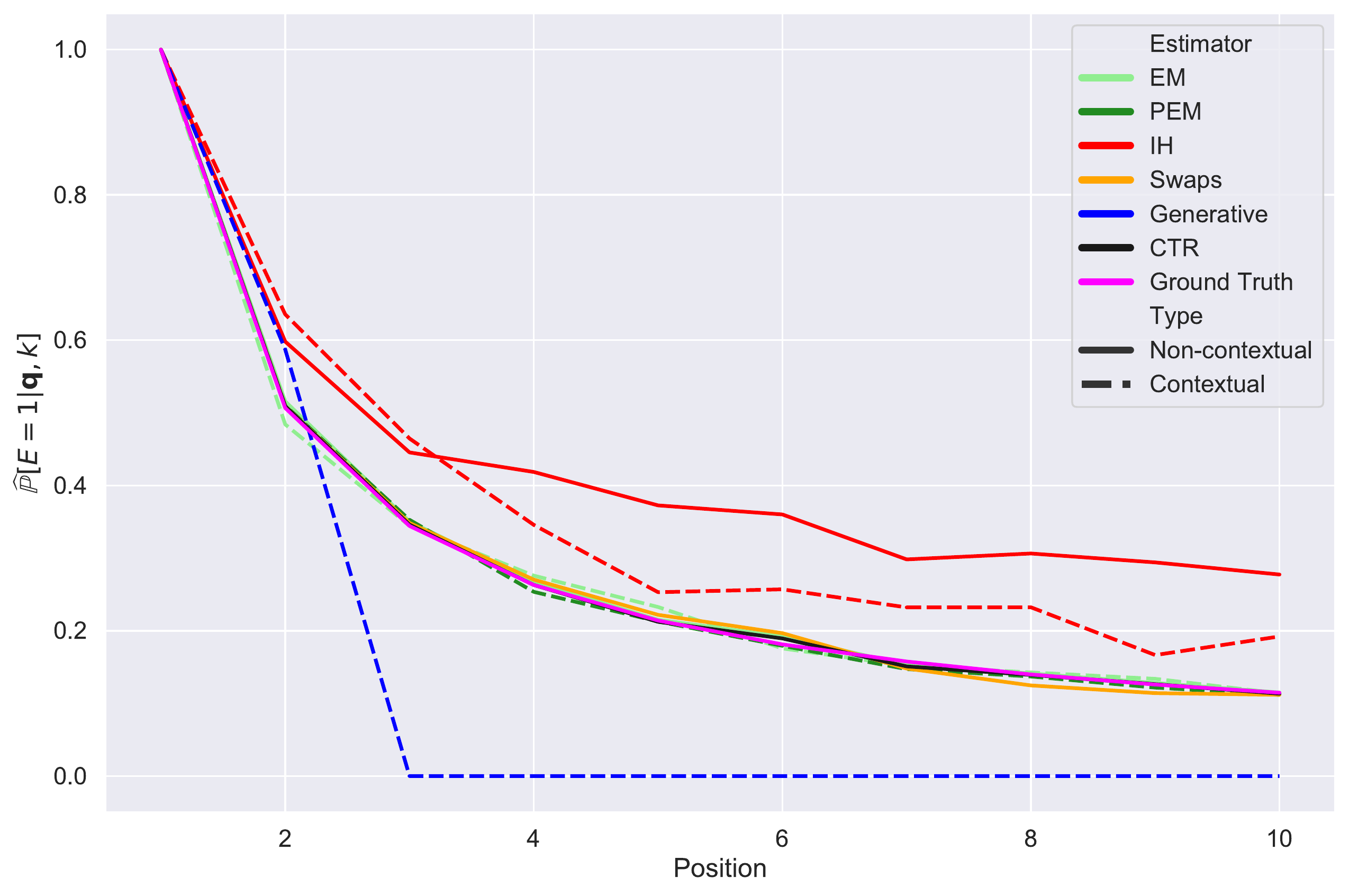}
	\caption{Average position bias over all the queries in the training set of the SINBIN dataset for $\eta=0.5$.}
	\Description{All estimators except IH-based and generative estimators can correctly fit the mean position bias.}
	\label{fig:meanpb-random}
\end{figure}

\subsection{Position bias estimation error}

In this section, we evaluate the quality of an estimator $f$ with the empirical version of the relative error \eqref{eq:relerror} as in \cite{contextual_ih}. Note that this equation needs the ground-truth contextual position bias, which is set to \eqref{eq:true_cpb} in all the experiments.

\begin{align}
	RelError(f)=\EEb{\qq}{\frac{1}{K}\sum_{k\in K}\abs{1-\frac{f(\qq, k)}{\prob{E=1|\qq,k}}}}
	\label{eq:relerror}
\end{align}

\subsubsection{Randomized rankings}\cref{fig:relerror-eta-sinbin} depicts the relative error for various estimators on the SINBIN dataset with different values of the context dependency parameter $\eta$. As expected, the non-contextual methods achieve a worst estimate of the position bias as $\eta$ increases. This is because the assumption that $E\independent \qq | k$ that the \gls*{pbm} follows is less realistic as $\eta$ grows. Recall that in non-contextual approaches, the estimated examination probabilities have no dependency on the context.

The contextual \gls*{em} methods consistently outperform all the other position bias estimators in this case. The latter achieve a lower relative error than the contextual generative model, even if both methods optimize the click likelihood under the \gls*{cpbm}.  This indicates the importance of applying the proposed \gls*{em}-based method for the estimation of the examination and relevance latent variables. The structure given by the marginals computed in the expectation step is proven crucial even when the maximization step is computed with the rough approximation given by a single mini-batch gradient update. Another interesting observation is the good performance of the \gls*{ctr} estimator. Even if this estimator makes a non-realistic assumption about the position bias, it offers the best position bias estimate among non-contextual methods along with the swap-based estimator. Given that the \gls*{ctr} estimator does not require randomization or information about the ranking process and only needs to perform a single pass over the data, it is worth taking it into account.

Given that the Yahoo \gls*{ltr} corpus has human ratings of the actions instead of computing them with linear regression, modeling the latent variable $R$ is potentially more complicated. We hypothesize that this causes the generative model to offer the poor performance depicted in \cref{fig:relerror-eta-yahoo}. As expected, the relative error of the non-contextual estimators show the same increasing trend with $\eta$. However, its value in this case is closer to that of the contextual methods. The non-contextual \gls*{ih} approach performs bests for $\eta \leq 0.5$ followed by the swap estimator. This is expected given that the non-contextual \gls*{ih} estimator has comparable performance to swap-based methods while it makes better use of the data \cite{unbiased_ltr}. The contextual \gls*{em} method has similar performance to the latter for $\eta\leq 0.5$ even if the information about the ranking randomization is not used. For $\eta>0.5$, nevertheless, the contextual \gls*{em} method consistently outperforms all the other estimators.

\begin{table}[t]
	\centering
	\begin{tabular}{rcc}
		\toprule
		Estimator & SINBIN & Contextual Yahoo \gls*{ltr} \\
		\midrule
		Contextual P\gls*{em} & \textbf{0.046} & 0.285 \\
		Contextual \gls*{em} & 0.054 & \textbf{0.241} \\
		Contextual \gls*{ih} & 0.411 & 0.665  \\
		Contextual Generative & 0.829& 4.179 \\
		\midrule
		Swaps & \textbf{0.318} & 0.191 \\
		CTR & 	0.325 & 0.430 \\
		\gls*{em} & 0.331 & 0.328\\
		\gls*{ih} & 1.092 &	\textbf{0.171} \\
		\bottomrule
	\end{tabular}
	\caption{Empirical relative error for $\eta=0.5$.}
	\label{tab:meanpb-random-relerror}
\end{table}

In \cref{fig:meanpb-random}, we see that the generative model and the \gls*{ih}-based methods fail at capturing the mean position bias across all queries for $\eta=0.5$ unlike the other methods. Note, however, that the position bias varies from one context to another given that $\eta\neq 0$. \Cref{tab:meanpb-random-relerror} presents the relative errors on the same experiment as that of \cref{fig:meanpb-random}. Successfully capturing the mean of the position biases is not enough to fit the position bias for each specific context. This is manifested in the difference of one order of magnitude between the proposed methods and the others for the SINBIN dataset. The proposed methods also outperform the other contextual estimators on the contextual Yahoo \gls*{ltr} dataset. In this case, however, the best estimations are obtained with the non-contextual \gls*{ih} and the swapping methods. Note that \cref{tab:meanpb-random-relerror} presents the results with $\eta=0.5$. As aforementioned, the latter methods are no longer the best estimators in terms of relative error as $\eta$ increases for $\eta>0.5$ (see \cref{fig:relerror-eta-yahoo}). Moreover, as discussed in \cref{sec:unbiased_ltr}, it does not provide the ranking boost performance characteristic of contextual methods.

\begin{figure}[t]
	\centering
	\includegraphics[width=.5\textwidth]{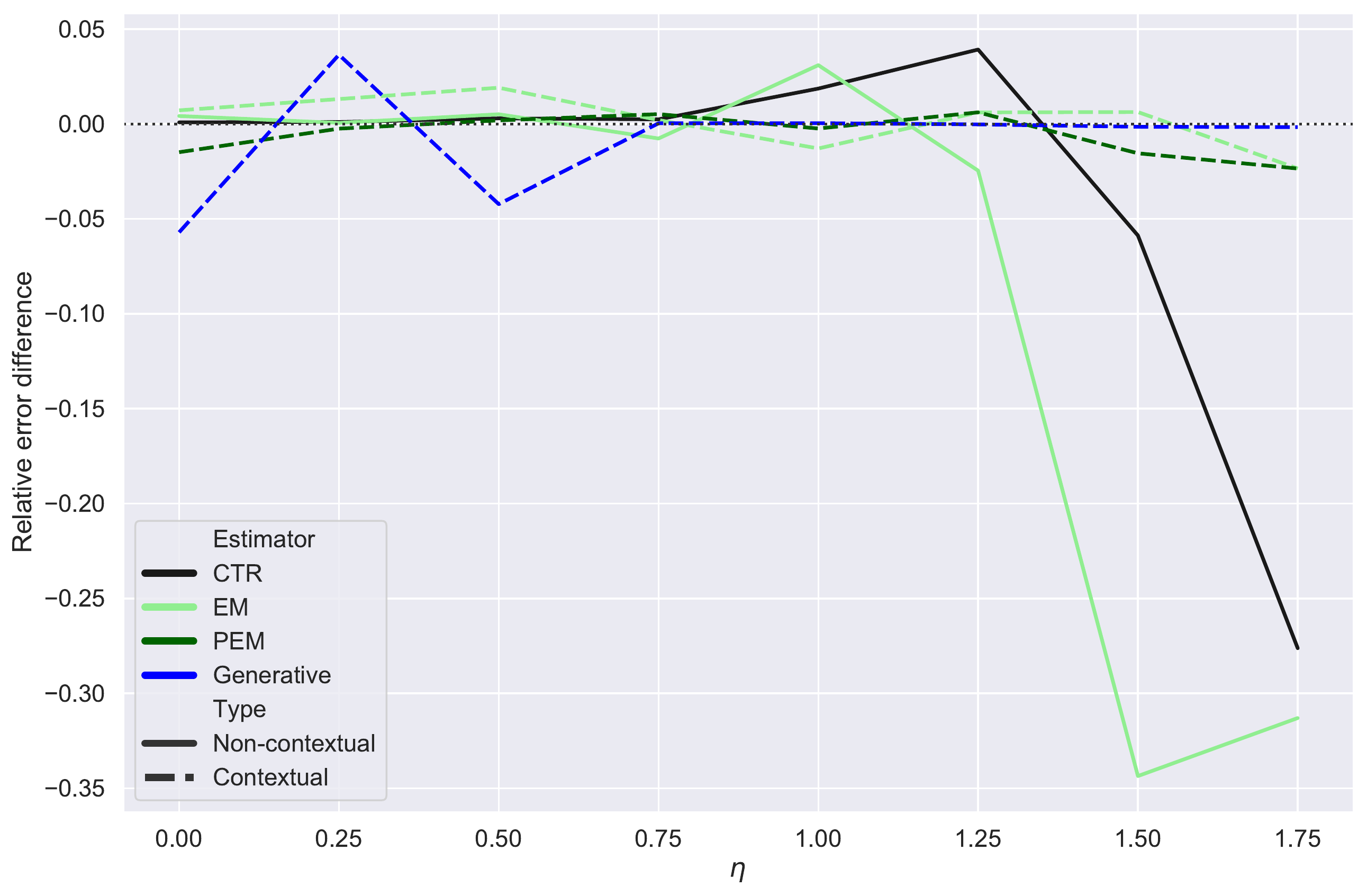}
	\caption{Difference in empirical relative error between the click log with and without randomized ranking for the SINBIN dataset. The dotted line indicates no difference. If the relative error difference is negative, the relative error is better when the ranking is randomized.}
	\Description{Randomization does affect the performance of the proposed methods. It helps EM and CTR methods for large eta, though.}
	\label{fig:relerrordiff-eta-sinbin}
\end{figure}

\subsubsection{Non-randomized rankings}In this experiment, we explore if methods that do not require ranking randomization benefit from contextual position bias estimation. The previous results have been obtained with implicit data collected after the randomization of the presented ranking. Such randomization is quantified, logged, and used by swap- and \gls*{ih}-based methods. When such randomization is not known or, in general, there is no information about the ranking process. Hence, we cannot use any of the former methods.

The other algorithms tested in this paper estimate the position bias without swapping the ranked items. However, some of them may benefit from such ranking randomization, as it promotes the clicks on lower ranked positions. \Cref{fig:relerrordiff-eta-sinbin} shows that this is the case for the non-contextual estimators on the SINBIN dataset: the position bias is more accurately estimated on the click log with randomized rankings for the \gls*{ctr} and non-contextual \gls*{em} methods for large $\eta$. 
We expect the relative error difference to be negative, especially for large $\eta$, where the non-contextual methods account for the largest relative error and the estimators benefit from more variety in the position of the clicks.

 \begin{figure}[t]
	\centering
	\includegraphics[width=.5\textwidth]{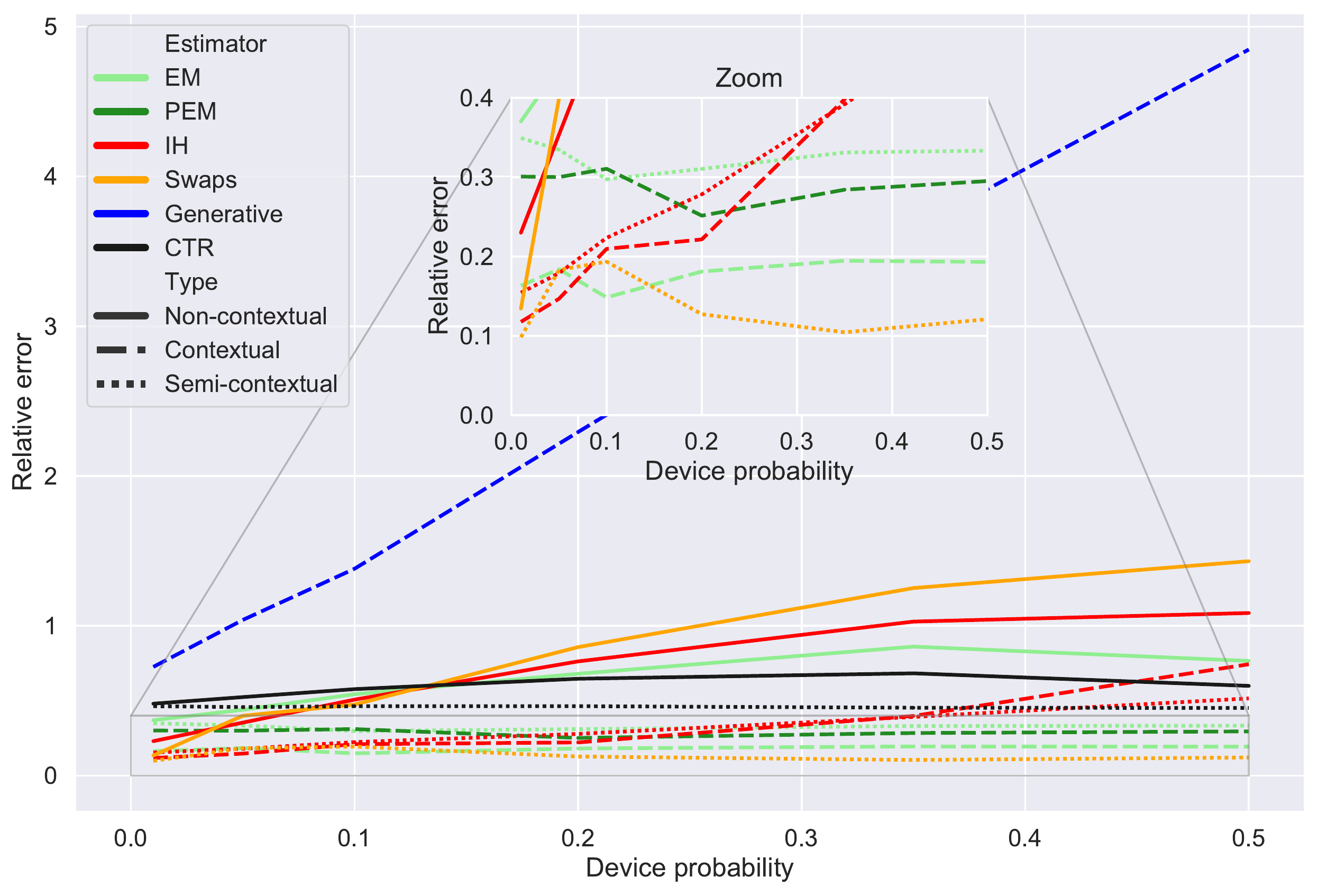}
	\caption{Empirical relative error as a function of the the device probability for the Contextual Yahoo \gls*{ltr} dataset with $\eta=1.5$.}
	\Description{Best method is the semi-contextual swap estimator. However, it needs randomization, and to correctly filter the queries by their difference in position bias distribution. Second best is the proposed method.}
	\label{fig:relerror-device-yahoo}
\end{figure}

\begin{figure*}[t]
	\centering
	\includegraphics[width=\textwidth]{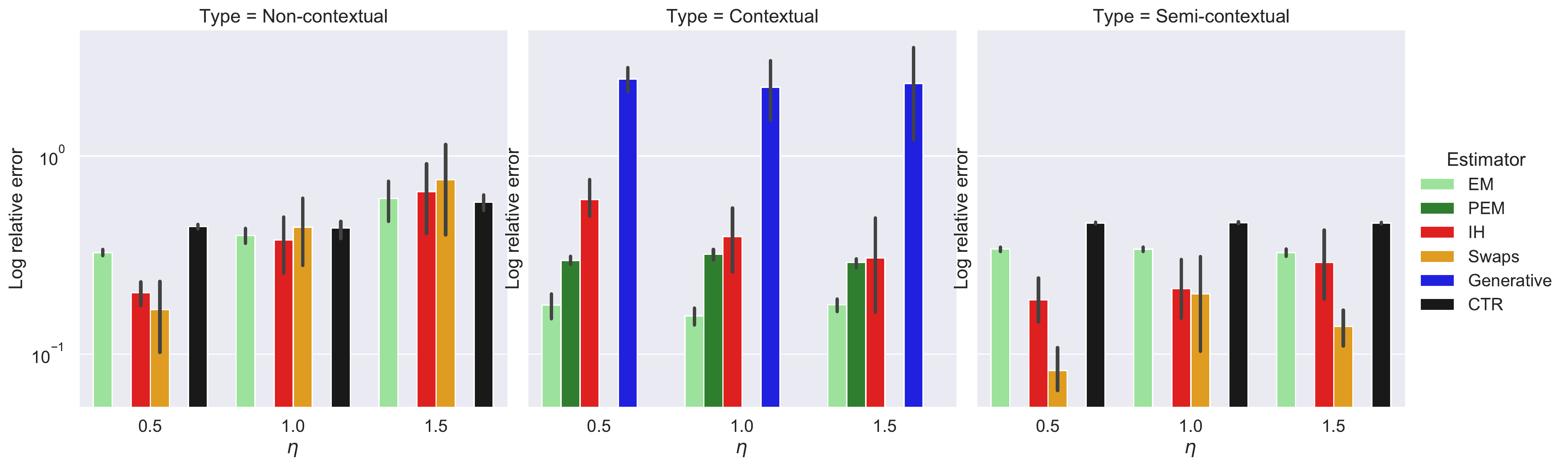}
	\caption{Empirical log-scaled relative error for different values of $\eta$ for the Contextual Yahoo \gls*{ltr} dataset. The variability of the values comes from the different values of device probability. The error bars represent the 95\% confidence intervals obtained with 1000 bootstrap iterations.}
	\Description{Swap-based semi-contextual approach is the best but has large variance for different device probabilities. Second best is the proposed EM estimator, which has low variance.}
	\label{fig:relerror-eta-device-yahoo}
\end{figure*}%
\begin{figure*}
	\begin{subfigure}[b]{\columnwidth}
		\centering
		\includegraphics[width=\textwidth]{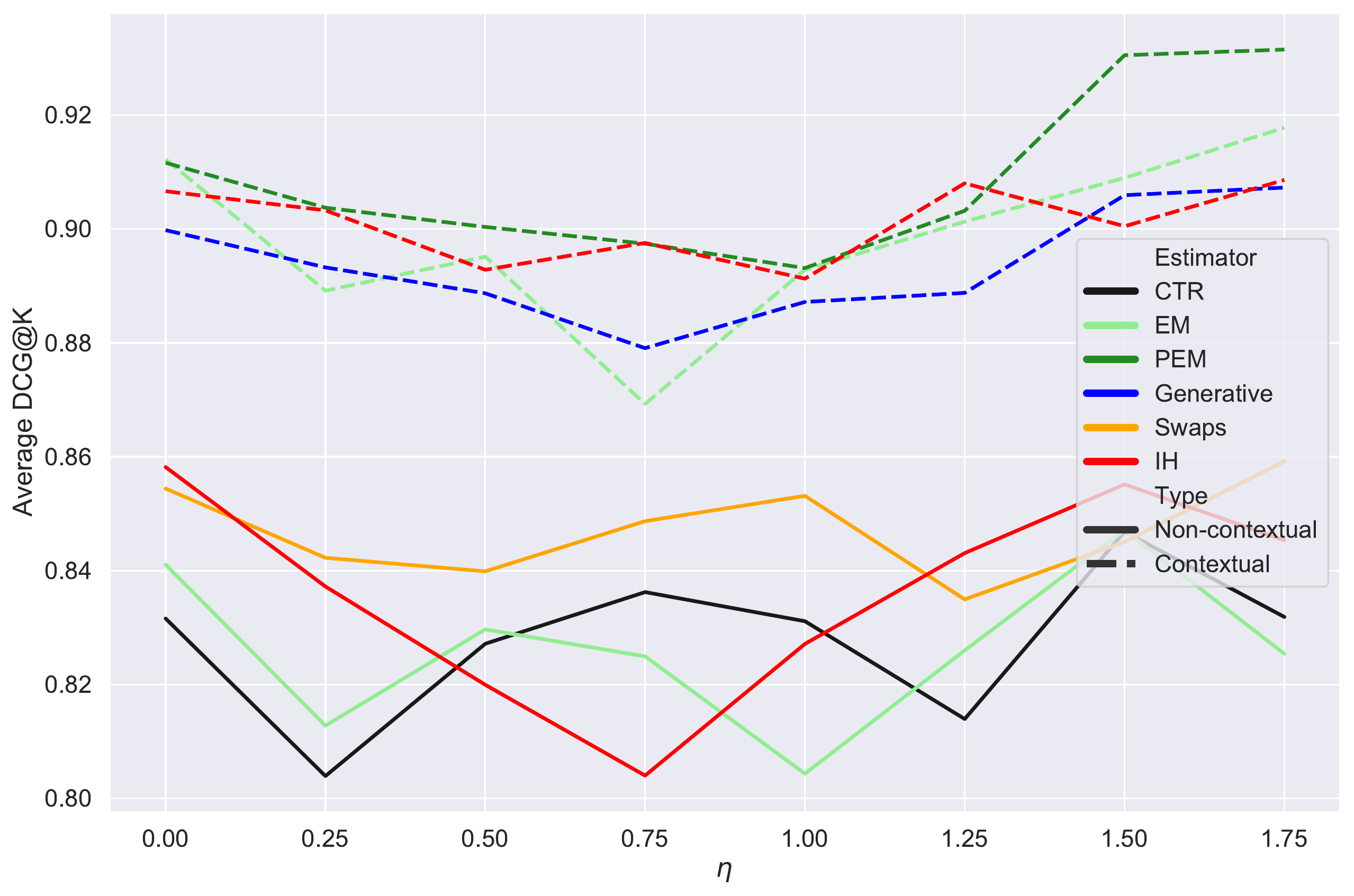}
		\caption{Average \gls*{dcg}@$K$}
		\Description{Ranking performance improves with respect to DCG at 10 with contextual estimates.}
		\label{fig:dcg10-eta-yahoo}
	\end{subfigure}%
	\begin{subfigure}[b]{\columnwidth}
		\centering
		\includegraphics[width=\textwidth]{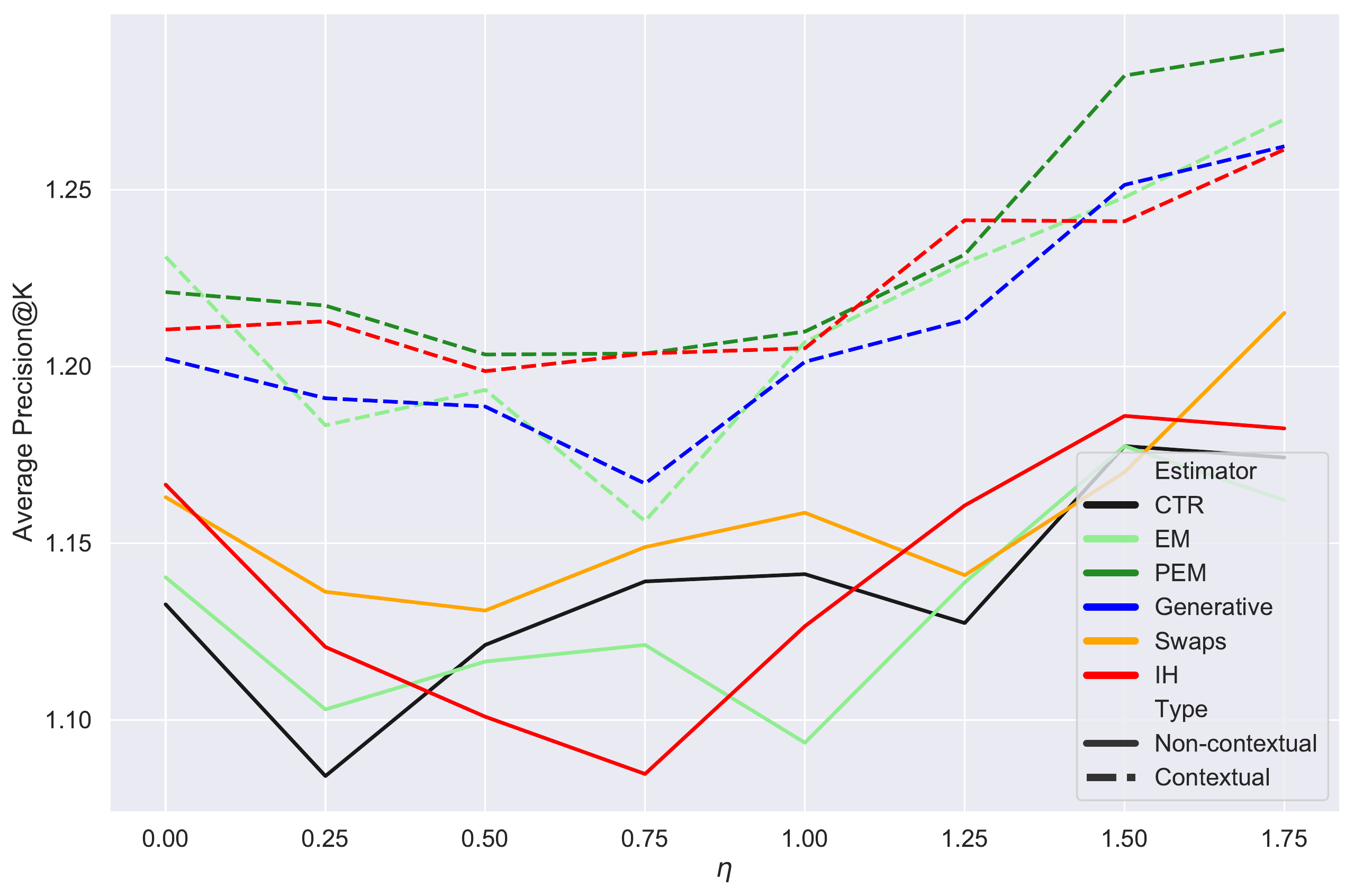}
		\caption{Average Precision@$K$}
		\Description{Ranking performance improves with respect to Precision at 10 with contextual estimates.}
		\label{fig:prec10-eta-yahoo}
	\end{subfigure}
	\caption{IR metric values as a function of the context importance $\eta$. These results have been obtained on the online \gls*{ltr} task on the test partition of the Contextual Yahoo \gls*{ltr} dataset.}
\end{figure*}

\subsubsection{Testing against semi-contextual approaches}As mentioned in \cref{sec:cpb_estimators}, semi-contextual approaches are powerful estimators when the \gls*{cpbm} holds. These estimators are suitable for a finite number of possibly different position bias curves and there exists a known correspondence between the different distributions of examination random variables and the values of the context vector features. One example of the latter is presented in \ref{it:queries}. Here, we have two types of queries, navigational and informational, each assigned with a different position bias. In this case, we can determine which one applies to a given query e.g., depending on the number of logical operators in the text query as in \cite{contextual_ih}.

In this experiment, we test the methods in \cref{sec:related_work,sec:methodology} in an ideal framework for semi-contextual approaches. We create two different synthetic devices and encode an identifier of the device being used for each query in the associated context vector. In particular, the device information is concatenated to the previously discussed context vector in the form of a one-hot encoding embedding. Therefore, in this experiment $\qq\in\R^{12}$, and the last two dimensions of the context vector are used to filter the data by device for the semi-contextual estimators. The contextual estimators, however, receive no information about the devices. That is, the contextual estimators use $\qq$ as in any other experiment. Since we want the position bias to only depend on the device type, we modify \eqref{eq:true_cpb} so that $\tilde{\ww}_i=0\ \forall i\in[10]$ and $\tilde{\ww}_i \overset{\text{i.i.d.}}{\sim} U(-\eta,\eta) \ \forall i \in\{11,12\}$. To sum up, in this case there are only two different position biases across all queries, and which one of them is used when generating the clicks for each query can be known with the last two dimensions of the vector $\qq$.

The device for each query is selected by sampling a Bernoulli random variable that we refer to as device probability. \Cref{fig:relerror-device-yahoo} shows the performance of various estimators for some values of such device probability. Note that we only plot the values in $[0,0.5]$ as the problem is symmetric. Similarly to the effect of $\eta$, augmenting the device probability makes the assumption of the \gls*{pbm} looser, thus yielding lower performance for non-contextual methods.

 As depicted in \cref{fig:relerror-device-yahoo}, the proposed \gls*{em} method achieves better performance than its non-contextual and semi-contextual versions. This proves the ability of the presented method to identify the different types of device and correctly model their position biases without prior knowledge. However, in this case, the best method is the semi-contextual swap-based estimator.
 
 The swap-based family of methods is statistically consistent as the number of samples grows \cite{ih}. Moreover, as shown in \cref{fig:relerror-eta-yahoo,tab:meanpb-random-relerror}, the swap estimator is also among the best estimators in practice when $\eta$ is small, that is when the \gls*{pbm} assumption is satisfied to some extent. Nevertheless, this estimator needs to filter the data by position bias type and present a randomized ranking to the user. The swap estimator has higher variance in the measured empirical relative error than the proposed methods for different values of device probability as shown in \cref{fig:relerror-eta-device-yahoo}. The contextual \gls*{em} estimator presents more stable values of empirical relative errors than the swap-based semi-contextual estimator across different values of both device probability and $\eta$. Finally, the contextual \gls*{em} estimator does not require ranking randomization.
 
\Cref{fig:relerror-eta-device-yahoo} reveals the increasing trend in the relative error on $\eta$ for non-contextual methods. Like in the other experiments the contextual \gls*{em}-based method dominates over the other contextual approaches and the the naive generative estimator performs poorly.

Overall, the contextual \gls*{em} method proposed in this paper offers the best estimates of the position bias, specially when the context dependency increases. In the toy example where the position bias only depends on the content type, the semi-contextual swap-based approach yields better estimates than our proposed method. However, the contextual \gls*{em} algorithm achieves the best performance among the other estimators, and is the most convenient approach for real scenarios where the type of position bias for each query is unknown. Our proposed algorithm does not require ranking randomization as the swapping approach, and consistently outperforms all the tested estimators when such randomization was either not used 
or not stored in the click log.

\subsection{Unbiased LTR performance}
\label{sec:unbiased_ltr}

The motivation to compute the position bias is to account for the unwanted confounders in the implicit data. The previous estimators can be used for \gls*{ltr} with biased feedback. In particular, given the correct estimation of the examination latent variable, the models learned on the debiased data will converge to those obtained with explicit feedback \cite{unbiased_ltr}. We investigate whether the context-dependent position bias estimates obtained using the \gls*{cpbm} offer a benefit over those following the \gls*{pbm}. To quantify the ranking performance improvement when data is unbiased, we evaluate the \textsc{LinTSPBMRank} ranker \cite{amazon_paper} with different position bias estimates. The ranker described in the original paper takes a context-independent estimate of the position bias, but the adaptation to use a context-dependent estimate is straightforward.

The \textsc{LinTSPBMRank} ranker is the same that is used when generating the click log. However, the results in this section are obtained on fresh data, with the previously stated position bias estimates to debias the data, and with the ranker randomly initialized. We train the ranker online and record some common IR metrics. 

\Cref{fig:dcg10-eta-yahoo} and \Cref{fig:prec10-eta-yahoo} depict the \gls*{dcg} and Precision at the last ranking position $K=10$ for different values of $\eta$, respectively. In both cases, there is a significant performance improvement when considering contextual position bias instead of its non-contextual version even when $\eta=0$.

Overall, the unbiased ranker that uses the contextual position bias computed with contextual P\gls*{em} offers the best performance for most values of $\eta$ with respect to the Precision and \gls*{dcg} metrics.

\section{Conclusions}
\label{sec:conclusions}

This paper proposed the contextual \gls*{em}-based regression method for position bias estimation. The presented algorithm provides a context-dependent estimation of the position bias that is computed using mini-batch updates, and can be used when the ranker(s) used on the click log are unknown. The ranker-agnostic nature of the estimator makes it preferable to the majority of position bias estimators, which rely on knowledge about the ranking process and, in most cases, on ranking randomization. The presented algorithm can be used on virtually any click log, and in the absence of ranking information on the data, it achieves better performance than its alternatives. To conclude, this paper introduced a convenient and versatile method for context-dependent position bias estimation, which is shown to greatly benefit the \gls*{ltr} algorithms learning on biased implicit data.

\balance
\bibliographystyle{ACM-Reference-Format}
\bibliography{references}


\end{document}